\documentclass[ps,prd,twocolumn,notitlepage,preprintnumbers,footinbib,superscriptaddress,longbibliography]{revtex4-1}

\usepackage{dcolumn}
\usepackage{bm}
\usepackage{amsmath}
\usepackage{amssymb}
\usepackage{color}
\usepackage{braket}
\usepackage{slashed}
\usepackage{xcolor}
\usepackage{graphicx}
\usepackage{mathrsfs}
\usepackage{subfig}
\usepackage{tikz}
\usepackage{comment}
\usepackage{cancel}
\usepackage{ulem}
\usepackage[pdfpagelabels=false,colorlinks=true,urlcolor=teal,citecolor=teal]{hyperref}
\usepackage[compat=1.1.0]{tikz-feynhand}
\allowdisplaybreaks
\newcommand{\bz}{\bar{z}}

\begin{document}

\title{
Approximate N$^2$LO and N$^3$LO QCD Predictions for $tW$ Production
} 

\author{Jia-Le Ding}
\affiliation{School of Physics, Shandong University, Jinan, Shandong 250100, China}

\author{Hai Tao Li}
\email{haitao.li@sdu.edu.cn}
\affiliation{School of Physics, Shandong University, Jinan, Shandong 250100, China}

\author{Jian Wang}
\email{j.wang@sdu.edu.cn}
\affiliation{School of Physics, Shandong University, Jinan, Shandong 250100, China}
\affiliation{Center for High Energy Physics, Peking University, Beijing 100871, China}

\begin{abstract}
We present high-precision predictions for associated 
$tW$ production at the LHC that incorporate the next-to-next-to-leading order hard and soft functions as well as the complete next-to-next-to-next-to-leading order scale-dependent terms derived from the corresponding anomalous dimensions. 
These higher-order corrections, which dominate the full perturbative results, increase the next-to-leading order cross section by more than 10\%. 
Based on the comparison with ATLAS and CMS measurements, we directly extract the Cabibbo-Kobayashi-Maskawa matrix element 
$|V_{tb}|=0.99\pm 0.03({\rm expt})\pm 0.03({\rm theor})$ 
without assuming unitarity, achieving a precision comparable with the current world average value.
\end{abstract}

\maketitle


The top quark is the heaviest elementary particle discovered so far.
Because of the large mass, it is widely accepted that the top quark has a close connection with new physics, 
e.g., supersymmetry \cite{Hall:1996gq},  that may appear only at high energy scales.
Search for this new physics demands a precise determination of the top-quark properties,
which can be achieved only through measurements of its production and decay.
At the Large Hadron Collider (LHC), the top-quark pair production, mainly induced by strong interaction, is the dominant channel 
and has been computed at QCD next-to-next-to-leading order (N$^2$LO) \cite{Czakon:2013goa,Catani:2019iny}.
However, the top-quark pair production is not sensitive to the electroweak coupling of the top quark.
The top quark decays almost exclusively into $Wb$ 
with the width predicted at next-to-next-to-next-to-leading order (N$^3$LO) \cite{Chen:2023dsi,Chen:2023osm}.
It can be used to determine the Cabibbo-Kobayashi-Maskawa (CKM) matrix element $V_{tb}$ assuming only unitarity of the matrix.
The single top-quark production, of which the cross section is around one-third of the pair production,
provides a direct probe of $V_{tb}$ without assuming unitarity.
Among the three single top-quark production channels, the $s$ and $t$ channels have been calculated at N$^2$LO \cite{Brucherseifer:2014ama,Berger:2016oht,Liu:2018gxa,Campbell:2020fhf}.
In contrast, the theoretical prediction of the $tW$ channel, i.e., $pp\to tW$, which has the second-largest production rate at the LHC,
still remains at next-to-leading order (NLO) \cite{Giele:1995kr,Zhu:2002uj,Campbell:2005bb,Cao:2008af},
a precision obtained over twenty-years ago.

There are two primary obstacles on the way to achieving a N$^2$LO prediction for $tW$ production.
The first obstacle is the challenging two-loop virtual correction that involves multiple scales in the loop diagrams.
Recently, two of the authors and collaborators have removed this obstacle by presenting 
 analytical and numerical results for the leading and full color contributions \cite{Chen:2022yni,Chen:2022pdw}, respectively.  
The other obstacle is a proper subtraction of the contribution from top-quark pair production,
which would interfere with the $tW$ production at higher orders.
Although the subtraction methods for the tree-level interference in the $ tW^-\bar{b}$ final state have been fully developed \cite{Belyaev:1998dn,Tait:1999cf,Campbell:2005bb,Frixione:2008yi}, 
the corresponding method at one-loop level was proposed only recently~\cite{Dong:2024jxp}
and the subtraction in the $ tW^-\bar{b}g$ final state is still missing.

In this work, we present high-precision predictions for $tW$ production
by utilizing the two-loop virtual correction and circumventing the second obstacle 
by focusing on the contribution in the soft radiation region.
One of the advantages of this method is that we do not need to consider the interference problem mentioned above,
because the invariant mass of a soft bottom quark and a $W$ boson cannot approach the top-quark mass shell.
Another advantage is that the soft radiation region 
not only captures the dominant contribution in the real corrections,
but also enables an efficient computation of the cross section to higher orders in perturbative theory due to its simple structure \cite{Kidonakis:2006bu,Kidonakis:2007ej,Kidonakis:2010ux,Kidonakis:2016sjf,Kidonakis:2019nqa,Li:2019dhg,Kidonakis:2021vob,Kidonakis:2024kts}.
As an essential building block, the threshold soft function that encompasses all effects of soft radiations has been calculated at ${\rm N^2LO}$ in our previous paper~\cite{Ding:2025xhc}.
In addition, by using the three-loop anomalous dimensions for the renormalization group (RG) evolution of the hard and soft functions~\cite{Liu:2022elt}, 
we can further derive all the large logarithmic (scale-dependent) terms up to ${\rm N^3LO}$.
All these  ingredients are coherently incorporated into our calculation framework
to give the  theoretical prediction for this process.

Comparison between our predictions and
the measurements of the ATLAS~\cite{ATLAS:2012bqt,ATLAS:2015igu,ATLAS:2016ofl,ATLAS:2020cwj,ATLAS:2024ppp} and CMS Collaborations~\cite{CMS:2012pxd,CMS:2014fut,CMS:2018amb,CMS:2021vqm,CMS:2022ytw,CMS:2024okz} at the LHC is of significant phenomenological interest.
The higher-order QCD corrections (beyond NLO) computed in this work increase the cross section obviously, 
leading to improved agreement between theoretical predictions and experimental measurements. 
Based on this comparison, we can determine the CKM matrix element $|V_{tb}|$ with an accuracy comparable to the world average value.

We consider the associated top quark and $W$ boson production at the LHC, i.e., $
    p(P_1)+p(P_2)\to t/\bar t(p_3)+W^-/W^+(p_4)+X(p_X),
$
where $X$ accounts for all additional radiation in the final state. 
 The Lorentz invariant kinematic variables  for this process are
  $s=(P_1+P_2)^2$ and $Q^2=(p_3+p_4)^2,$
which define the ratio $\tau=Q^2/s$.
In the soft radiation region, the energy component of the four-momentum $p_X^{\mu}$ is highly suppressed, i.e.,
$p_X^0=\sqrt{s}(1-\tau)/2\to 0$, which is called the hadronic threshold limit.
For comparison, the threshold variable in Refs. \cite{Kidonakis:2007ej,Kidonakis:2010ux,Kidonakis:2016sjf,Kidonakis:2019nqa,Kidonakis:2021vob} is defined by 
$S_4=2p_4\cdot p_X$, which measures the soft limit in a different way.

\begin{figure}
    \centering
    \begin{tikzpicture}[scale=0.5]
        \begin{feynhand}
            \vertex [particle] (a) at (0,0) {\large $b$};
            \vertex [particle] (b) at (0,3) {\large $g$};
            \vertex [particle] (c) at (5,0) {\large $W^-$};
            \vertex [particle] (d) at (5,3) {\large $t$};
            \vertex (e) at (1.5,1.5);
            \vertex (f) at (3.5,1.5);
            \propag [fer] (a) to (e);
            \propag [fer] (e) to (f);
            \propag [fer] (f) to (d);
            \propag [pho] (f) to (c);
            \propag [glu] (b) to (e);
        \end{feynhand}
    \end{tikzpicture}
    \quad
    \begin{tikzpicture}[scale=0.5]
        \begin{feynhand}
            \vertex [particle] (a) at (0,0) {\large $b$};
            \vertex [particle] (b) at (0,3) {\large $g$};
            \vertex [particle] (c) at (5,0) {\large $W^-$};
            \vertex [particle] (d) at (5,3) {\large $t$};
            \vertex (f) at (2.5,0);
            \vertex (e) at (2.5,3);
            \propag [fer] (a) to (f);
            \propag [fer] (f) to (e);
            \propag [fer] (e) to (d);
            \propag [pho] (f) to (c);
            \propag [glu] (b) to (e);
        \end{feynhand}
    \end{tikzpicture}
    \caption{LO Feynman diagrams for $tW^-$ production.}
    \label{fig:feynman diagrams}
\end{figure}

The corresponding LO partonic scattering process is $
    b/\bar b(p_1)+g(p_2)\to t/\bar t(p_3)+W^-/W^+(p_4)
$
with the LO Feynman diagrams shown in Fig. \ref{fig:feynman diagrams}.
The partonic center-of-mass energy $\sqrt{\hat s}$ is defined by $\hat{s}=(p_1+p_2)^2$,
which can be used to define the variable $z=Q^2/\hat{s}$.
The partonic threshold is described by $\bar{z}=1-z\to 0$.

The soft radiation induces large logarithmic terms in the cross section, which dominate higher-order QCD corrections. 
In the threshold limit,  the differential hadronic cross section can be expressed in a factorization form 
\begin{align}
   \frac{{d\sigma }}{{d{Q^2}d{\Phi _2}}} = \frac{1}{s}\int_{\tau}^1 {\frac{{dz}}{z}} \mathcal{L}\left( {\frac{{\tau}}{{z}},\mu } \right)&\frac{1}{{2Q^2}}
   H\left( {\beta_t , y,\mu} \right)\nonumber\\
   &\times\mathcal{S}\left( \bar{z} ,\beta_t ,y ,\mu\right),\label{eq:dsigma}
\end{align}
where $\Phi_2$ denotes the two-body phase space of the top quark and the $W$ boson, and the parton luminosity $\mathscr{L}$ is defined as the convolution
\begin{align}
    \mathcal{L}\left( {\frac{{\tau}}{{z}},\mu } \right) = \int_{{\tau}/ {z} }^1 {\frac{{d{x}}}{{{x}}}} \Bigg[&f_b\left( {{x},\mu } \right)f_g\left( {\frac{{\tau}}{{{x}z}},\mu } \right)+(b\leftrightarrow g) 
    \Bigg]\nonumber
\end{align}
with $f_{b/g}(x,\mu)$ being the parton distribution function (PDF) at the scale $\mu$.
The PDF is strongly suppressed in the region $x \to 1$.
Therefore, the parton luminosity is dynamically enhanced as $z\to1$;
i.e.,  the partonic threshold limit can be reached even if the hadronic limit $\tau \to 1$ is not fully realized~\cite{Becher:2007ty}.   
The hard function $H$ is process dependent and can be obtained by calculating the corresponding virtual correction 
to the partonic scattering process.
It is a function of $y\equiv\cos\theta$ with $\theta$  the polar angle of the outgoing top quark relative to the incoming bottom quark and
the top-quark velocity,  which is defined by
  $  \beta_t=\sqrt{(Q^2-m_t^2-m_W^2)^2-4m_t^2m_W^2}/(Q^2+m_t^2-m_W^2).$
The soft function $\mathcal{S}$ encodes the effects of soft radiations 
in terms of logarithms $((\ln\bar{z})^{2n-i}/\bar{z})_+$ with $i=1,\cdots,2n$ at the order of $\alpha_s^n$, 
and $\delta(1-z)$ terms.
The subscript `+' indicates that the pole at $z=1$ is subtracted, e.g.,
\begin{align}
    \int_0^1 dz g(z) \left( \frac{\ln\bar{z}}{\bar{z}}\right)_+  =     \int_0^1 dz [g(z)-g(1)]  \frac{\ln\bar{z}}{\bar{z}}\,
 \end{align}
with $g(z)$ being a regular function.

Since the large logarithms have already provided dominant contributions to the cross section,
a precise knowledge of their structure at higher orders is valuable for phenomenological studies, 
when obtaining the full perturbative prediction remains a formidable challenge.
From the effective field point of view, 
the logarithms arise because there are different scales in the process.
In our case, they are the hard scale $\sqrt{\hat{s}}$ and the soft scale $\sqrt{\hat{s}}-Q$.
If the hard and soft functions are calculated at their intrinsic scales, 
no large logarithms would appear.
Choosing a common scale, such as  $\mu$ in Eq. (\ref{eq:dsigma}), 
leads to the logarithms either in the hard function or the soft function.
This understanding can be used to develop a resummation scheme 
with the logarithms predicted to all orders, i.e., the hard (soft) function is calculated at the hard (soft) scale and evolved to other scales using the RG equation.
Explicitly, the resummation improved cross section can be written as 
\begin{align}
   &\frac{{d\sigma }}{{d{Q^2}d{\Phi _2}}} = \frac{1}{s}\int_{\tau}^1 {\frac{{dz}}{z}} \mathcal{L}\left( {\frac{\tau}{z},\mu_f } \right)\frac{1}{{2Q^2}}H\left( \beta_t , y,{\mu_h} \right)
  \nonumber\\
   &\times U(\mu_h,\mu_s, \mu_f)\tilde{\mathcal{S}}\left(\partial_\eta,\beta_t ,y \right)\frac{1}{\bz}\left(\frac{\bz^2 Q^2}{z\mu^2_s}\right)^{\eta}
   \frac{e^{-2\gamma_E\eta}}{\Gamma(2\eta)}
   \label{eq:RGtot}    
\end{align}
where the evolution factors $U$ and $\eta$ depend on the hard and soft scales, 
denoted by $\mu_h$ and $\mu_s$, respectively, and the factorization scale $\mu_f$.
If we choose $\mu_h$ and $\mu_s$ at $\sqrt{\hat{s}}$ and $\sqrt{\hat{s}}-Q$, respectively,
all the logarithms are encoded in the evolution factors.
On the other hand, if we set all the scales equal, i.e., $\mu_h=\mu_s=\mu_f=\mu$,
no evolution effects exist, and thus $U=1$ and $\eta=0$.
Note that the Laplace transformed soft function $\tilde{\mathcal{S}}$ depends on the derivative $\partial_{\eta}$.
Its action on the following part must be taken before setting $\eta=0$.
Both the hard and soft functions have been calculated perturbatively up to N$^2$LO in Refs.~\cite{Chen:2022yni,Chen:2022pdw} and Ref.~\cite{Ding:2025xhc}, respectively.
Below we derive the ${\rm N^3LO}$ results for hard and soft functions  by solving the RG evolution equations.

The RG evolution of the Laplace transformed soft function $\tilde{\mathcal{S}}(\ln(\hat{s}\xi^2/\mu^2),\beta_t,y)\equiv \int_0^{\infty} d\bar{z} e^{-\bar{z}/(e^{\gamma_E}\xi)}\mathcal{S}(\bar{z},\beta_t,y,\mu)$ reads
\begin{align}
    \frac{{d\tilde{\mathcal{S}}}(L_{\mu},\beta_t,y)}{{d\ln \mu }} = {\gamma _s}\tilde{\mathcal{S}}(L_{\mu},\beta_t,y),
    \label{eq:rgsoft}
\end{align}
where $L_{\mu}=\ln(\hat{s}\xi^2/\mu^2)$ and the soft anomalous dimension is given by
\begin{align}
    &{\gamma _s} =  - \left( {{\mathbf{T}}_1^2 + \mathbf{T}_2^2} \right){\gamma _{{\rm{cusp}}}}L_{\mu}- 2{\gamma_q} - 2{\gamma_g} - 2{\gamma_Q} \nonumber\\
    &+ \mathbf{T} _1\cdot \mathbf{T}_3{\gamma _{{\rm{cusp}}}}\ln \frac{{{{\left( {1 - \beta_t y} \right)}^2}}}{{1 - {\beta_t ^2}}}+ \mathbf{T}_2 \cdot \mathbf{T}_3{\gamma _{{\rm{cusp}}}}\ln \frac{{{{\left( {1 + \beta_t y} \right)}^2}}}{{1 - {\beta_t ^2}}}\nonumber\\
    & - {\rm{2}}\gamma _f^q - {\rm{2}}\gamma _f^g- 2{{\cal T}}_{1233}\Big(\frac{\alpha_s}{4 \pi}\Big )^3{\mathcal{F}_{h2}}\left( {\frac{{1 - {\beta_t ^2}}}{{1 - {\beta_t ^2}{y^2}}}} \right)\,.
  \label{eq:softano}  
\end{align}
Here $\mathbf{T}_i$ represents the color charge of the $i$-th parton \cite{Catani:1996jh,Catani:1996vz}. In the tripole correlation, $\mathcal{T}_{1233}$ is a new color structure  that first appears at three-loop level, defined as
\begin{align}
    \mathcal{T}_{1233}=\frac{1}{2}f^{ade}f^{bce}\mathbf{T}_1^a\mathbf{T}_2^b(\mathbf{T}_3^c\mathbf{T}_3^d+\mathbf{T}_3^d\mathbf{T}_3^c),
\end{align} 
and for $tW$ production $\mathcal{T}_{1233}$ can be replaced by $-3/4$. 
The associating function $\mathcal{F}_{h2}$ was calculated in \cite{Liu:2022elt}. 
The cusp $\gamma_{\rm cusp}$ and the other anomalous dimensions $\gamma_q,\gamma_g,\gamma_Q$ have been obtained up to three loops~\cite{Moch:2004pa,Becher:2009qa,Bruser:2019yjk},
i.e., the first three terms in the expansion $\gamma =  \sum_{n=0} (\alpha_s/4\pi)^{n+1} \gamma^{(n)}$ are known.
Solving Eq. (\ref{eq:rgsoft}), we obtain 
\begin{align}
\tilde{\mathcal{S}}(L_{\mu},\beta_t,y) = \tilde{\mathcal{S}}(0,\beta_t,y) \exp\left(\int_{\ln\mu_s}^{\ln\mu}\gamma_s d\ln \mu\right)
\label{eq:solvesoft}
\end{align}
with $\mu_s=\sqrt{\hat{s}}\xi$.
Expansion of the above expression gives the N$^3$LO soft function,
in which the scale-dependent terms are fully determined by the exponential function and thus exact.
The only missing ingredient is the scale-independent part 
$\tilde{\mathcal{S}}^{(3)}(0,\beta_t,y)$ under the notation $\tilde{\mathcal{S}}=\sum_{n=0}(\alpha_s/4\pi)^n\tilde{\mathcal{S}}^{(n)}$.

The RG evolution of the Laplace transformed PDFs  in the threshold limit  can be found in Refs.~\cite{Korchemsky:1992xv,Moch:2004pa}, 
\begin{align}
    \frac{d\tilde{f}_i(\xi,\mu)}{d\ln\mu} =(2\mathbf{T}_i\cdot\mathbf{T}_i \gamma_{\text {cusp}} \ln \xi+2\gamma_{f}^i)\tilde f_i(\xi,\mu)
\end{align}
with the anomalous dimensions  $\gamma_f^i~(i=q,g)$ well-known \cite{Becher:2007ty,Ahrens:2009cxz}. 
Consequently, the RG evolution of the hard function can be derived from the fact that the total physical cross section must be independent of the  scale $\mu$, 
expressed as
\begin{align}
    \frac{d\ln H}{d\ln\mu} =- \frac{d \ln{\tilde f}_q}{d\ln\mu}- \frac{d\ln {\tilde f}_g }{d\ln\mu} - \frac{d\ln\tilde{\mathcal{S}}}{d\ln\mu},
\end{align}
which can be solved similarly to Eq. (\ref{eq:solvesoft}).

Inserting the perturbative hard and soft functions into Eq. (\ref{eq:RGtot})
and expanding 
\begin{align}
    \frac{1}{\bz}\left(\frac{\bz^2Q^2}{z\mu^2}\right)^{\eta}=
  B(1-\eta,2\eta)  
    \left(\frac{Q^2}{\mu^2}\right)^{\eta}\delta(\bz)+\left[\frac{1}{\bz}\left(\frac{\bz^2Q^2}{z\mu^2}\right)^{\eta}\right]_+
    \nonumber
\end{align}
we derive the fixed-order partonic cross section in the threshold limit that contains a series of logarithmically enhanced terms, 
\begin{align} \label{eq:an3lo}
    &\frac{d\hat \sigma}{dQ^2d\Phi_2}=\frac{d\hat \sigma_{\text{LO}}}{dQ^2d\Phi_2} 
    \sum_{n=0}\bigg(\frac{\alpha_s(\mu_r)}{4\pi}\bigg)^n \Bigg(\sum_{m=-1}^{2n-1}C_{n,m}P_m\Bigg)
   \nonumber
\end{align}
where $d\hat{\sigma}_{\rm LO}$ is the LO differential cross section and
 $P_m$ denotes the distribution $\Big[\frac{1}{\bz}\left(\ln\frac{\bz^2 Q^2}{z\mu_f^2}\right)^m \Big]_+$ for $m\ge 0$ and $\delta(1-z)$ for $m=-1$.
The corresponding coefficients are given by
\begin{align}
    &C_{n,2n-1}=\frac{n}{4^{n-1} n!}(-\gamma_{s,d}^{(0)})^n,\nonumber\\
    &C_{n,2n-2}=-\frac{n(2n-1)}{4^{n-1} n!}(-\gamma_{s,d}^{(0)})^{n-1}\Bigg[\gamma_{s,c}^{(0)}+\frac{2(n-1)}{3}\beta_0\Bigg],\nonumber\\
    &C_{1,-1}=-\frac{1}{4}\ln^2\frac{Q^2}{\mu_f^2}\bigg(\gamma_{h,d}^{(0)}+\gamma_{s,d}^{(0)}\bigg)-\frac{1}{2}\ln\frac{Q^2}{\mu_f^2}\bigg(\gamma_{h,c}^{(0)}+\gamma_{s,c}^{(0)}\bigg)\nonumber\\
    &-\beta_0\ln\frac{Q^2}{\mu_r^2}+H_c^{(1)}+\tilde{\mathcal{S}}_c^{(1)},\nonumber\\
    &C_{2,1}=-\frac{\pi^2}{2}(\gamma_{s,d}^{(0)})^2+\frac{1}{4}\Bigg[\ln^2\frac{Q^2}{\mu_f^2}\gamma_{s,d}^{(0)}\gamma_{h,d}^{(0)}\nonumber\\
    &+2\ln\frac{Q^2}{\mu_f^2}\gamma_{s,d}^{(0)}\bigg(\gamma_{h,c}^{(0)}-2
    \beta_0\bigg)+8\beta_0\ln\frac{Q^2}{\mu_r^2}\gamma_{s,d}^{(0)}\nonumber\\
    &+2\gamma_{s,c}^{(0)}\Big(\gamma_{s,c}^{(0)}+2\beta_0\Big)-4\gamma_{s,d}^{(0)}\Big(H_c^{(1)}+\tilde{\mathcal{S}}_c^{(1)}\Big)-4\gamma_{s,d}^{(1)}\Bigg],\nonumber\\
    &C_{3,3}=\frac{5\pi^2}{24}(\gamma_{s,d}^{(0)})^3-\frac{\gamma_{s,d}^{(0)}}{48}\Bigg[3\ln^2\frac{Q^2}{\mu_f^2}\gamma_{s,d}^{(0)}\gamma_{h,d}^{(0)}\nonumber\\
    &+6\ln\frac{Q^2}{\mu_f^2}\gamma_{s,d}^{(0)}\bigg(\gamma_{h,c}^{(0)}-4\beta_0\bigg)+36\beta_0\ln\frac{Q^2}{\mu_r^2}\gamma_{s,d}^{(0)}\nonumber\\
    &+4\gamma_{s,c}^{(0)}\Big(3\gamma_{s,c}^{(0)}+10\beta_0\Big)-12\gamma_{s,d}^{(0)}\Big(H_c^{(1)}+\tilde{\mathcal{S}}_c^{(1)}\Big)\nonumber\\
    &+16\beta_0^2-24\gamma_{s,d}^{(1)}\Bigg],
\end{align}
where the anomalous dimensions are defined by
\begin{align}
    \gamma_{s,d}&=-\gamma_{\text{cusp}}({\rm C_A+C_F}),\nonumber\\
    \gamma_{s,c}&=\gamma_{\text{cusp}}\Big[\frac{{\rm C_A-2C_F}}{2}\ln\frac{(1-\beta_t y)^2}{1-\beta_t^2}
    -\frac{\rm C_A}{2}\ln\frac{(1+\beta_t y)^2}{1-\beta_t^2}\Big]\nonumber\\
    &-2\gamma_f^{q}-2\gamma_f^{g}-2\gamma_q-2\gamma_g-2\gamma_Q, \nonumber\\
    \gamma_{h,d}&=\gamma_{\text{cusp}}({\rm C_A+C_F}),\nonumber\\
    \gamma_{h,c}&=\gamma_{\text{cusp}}\Big[\frac{\rm C_A}{2}\ln\frac{(1+\beta_t y)^2}{1-\beta_t^2}-\frac{{\rm C_A-2C_F}}{2}\ln\frac{(1-\beta_t y)^2}{1-\beta_t^2}\Big]\nonumber\\
    &+2\gamma_q+2\gamma_g+2\gamma_Q,
\end{align}
and $\beta_0=11\mathrm{C_A}/3-4\mathrm{T_F}n_f/3$ with $n_f$ being the number of massless quark flavors 
and the color factors $C_A=3$ and $C_F=4/3$ in QCD.
We have used $H_c$ and $\tilde{\mathcal{S}}_c$ to represent the scale-independent parts of hard and soft functions, respectively.

\renewcommand{\arraystretch}{1.5}
\begin{table*}[ht]
    \centering
    \begin{tabular}{|c|c|c|c|c|c|}
    \hline
    $\sqrt{s}$ & 7 TeV & 8 TeV& 13 TeV& 13.6 TeV& 14 TeV \\ \hline
    LO & $10.7_{-1.8}^{+2.3}\pm0.3$ & $15.4_{-2.5}^{+3.2}\pm0.4$ & $51.4_{-7.3}^{+9.3}\pm1.1$ & $57.1_{-8.1}^{+10.2}\pm1.2$ & $60.9_{-8.6}^{+10.8}\pm1.2$ \\ \hline
    NLO & $14.6_{-0.5}^{+0.4}\pm0.5$ & $20.9_{-0.7}^{+0.6}\pm0.7$ & $68.7_{-2.9}^{+2.5}\pm1.4$ & $76.1_{-3.2}^{+2.7}\pm1.5$ & $81.3_{-3.5}^{+2.8}\pm1.6$ \\ \hline
    ${\rm N^2LO_{LP}}$ & $18.5_{-0.8}^{+0.6}\pm0.5$ & $26.3_{-1.0}^{+0.8}\pm0.7$ & $84.8_{-2.8}^{+2.3}\pm1.7$ & $93.7_{-3.1}^{+2.5}\pm1.8$ & $99.9_{-3.1}^{+2.6}\pm1.9$ \\ \hline
    ${\rm N^3LO_{LP}}$ & $19.1_{-0.3}^{+0.2}\pm0.5$ & $27.1_{-0.4}^{+0.2}\pm0.6$ & $86.8_{-1.0}^{+0.5}\pm1.7$ & $95.9_{-1.1}^{+0.6}\pm1.9$ & $102.2_{-1.2}^{+0.6}\pm1.9$ \\ \hline
    ${\rm aN^2LO}$ & $16.6_{-0.5}^{+0.4}\pm0.5$ & $23.8_{-0.7}^{+0.6}\pm0.6$ & $77.2_{-2.2}^{+1.8}\pm1.6$ & $85.4_{-2.5}^{+2.1}\pm1.7$ & $91.2_{-2.8}^{+2.1}\pm1.8$ \\ \hline
    ${\rm aN^3LO}$ & $17.2_{-0.9}^{+0.8}\pm0.5$ & $24.5_{-1.3}^{+1.1}\pm0.7$ & $79.2_{-3.8}^{+3.5}\pm1.6$ & $87.6_{-4.2}^{+3.9}\pm1.7$ & $93.5_{-4.6}^{+4.1}\pm1.8$ \\ \hline
    \end{tabular}
    \caption{The fixed-order predictions for total cross sections of $tW$ production (in pb). The first and second errors represent the scale and PDF uncertainties, respectively.}\label{tab:fixorder}
\end{table*}

Note that the renormalization and factorization scales, $\mu_r$  and $\mu_f$, have been  clearly distinguished.
This is made possible by making use of the fact that the partonic cross section is independent of the renormalization scale order-by-order in $\alpha_s$ and the dependence of $\alpha_s(\mu_r)$ on $\mu_r$ is predictable.

In the above equations, we have presented the results for the first few leading logarithms explicitly.
The coefficients of the other subleading logarithms are too lengthy to be shown here and can be found in Supplemental Material.
All coefficients are exact except for $C_{3,-1}$, in which the three-loop scale-independent parts of the hard and soft functions, $H_c^{(3)}$ and $\tilde{\mathcal{S}}_c^{(3)}$, are still lacking.

The above expression can be considered as the leading power (LP) expansion in $\bz$, i.e., the $\mathcal{O}(\bz^{-1})$ contribution in the cross section, denoted by $\textrm {N}^{n}\textrm{LO}_{\rm LP}$.
The subleading power corrections consist of the higher power terms in the $gb$ channel and the contribution of the other partonic channels,
both of  which are of $\mathcal{O}(\bz^{i})$ with $i\ge 0$.
These corrections do not have a general structure that can be predicted without 
a full calculation
and also suffer from the complicated problem of interference with  $t\bar{t}$ production.
We include them in QCD NLO theoretical predictions and adopt the power subtraction to deal with the interference problem \cite{Dong:2024jxp}.
As such, the approximate N$^n$LO ($n\ge 2$) prediction of the cross section is defined as
\begin{align}\label{eq:NNNLO}
    d\sigma(\textrm {aN}^{n}\textrm{LO}) 
     = d\sigma(\textrm {N}^{n}\textrm{LO}_{\rm LP}) +  d\sigma({\rm NLO})-d\sigma({\rm NLO_{LP}}) .
     \nonumber
\end{align}

Our calculation framework is different from the previous study based on the traditional resummation formalism \cite{Kidonakis:2007ej,Kidonakis:2010ux,Kidonakis:2016sjf,Kidonakis:2019nqa,Kidonakis:2021vob,Kidonakis:2024kts} in several aspects.
First, the threshold variable, which depicts how the soft limit is approached, is defined differently, 
as mentioned above.
Second, the ${\rm N^2LO}$ scale-independent parts of the hard and soft functions, which would also enter the ${\rm N^3LO}$ scale dependent terms, are not included in previous computations.
Third, the formalism in Refs. \cite{Kidonakis:2021vob,Kidonakis:2024kts} employs the three-loop soft anomalous dimensions \cite{Kidonakis:2019nqa} that do not take into account the tripole correlation involving one massive and two massless partons.
The numerical difference is discussed below.

\begin{figure}[h]
    \centering
    \includegraphics[width=0.98\linewidth]{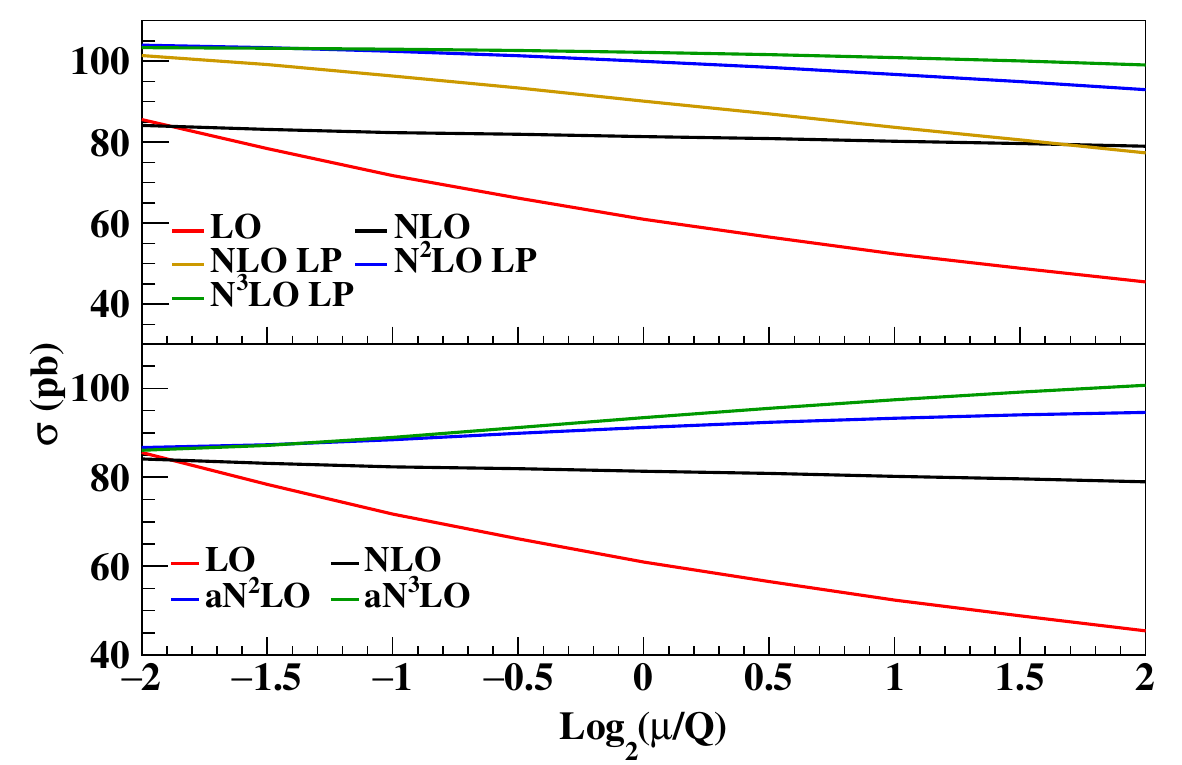}
    \caption{Scale dependence of the inclusive $tW$ cross section with $\mu_r=\mu_f=\mu$ at the 14 TeV LHC. 
    }
    \label{fig:scales}
\end{figure}

For the numerical evaluation, we take the electroweak parameters,
$  m_t =172.57~{\rm GeV}, m_W  =80.369~{\rm GeV},  G_F=1.16639\times10^{-5}~{\rm GeV^{-2}}
$ \cite{ParticleDataGroup:2024cfk},
and utilize the {\tt PDF4LHC21} PDF sets~\cite{PDF4LHC:2022cjn} along with the associated $\alpha_s(m_Z)=0.118$ in the LHAPDF library~\cite{Buckley:2014ana}. 
The theoretical uncertainties of our predictions are estimated from two primary sources.
The scale uncertainty, which evaluates the magnitude of the missing higher-power and higher-order corrections, is derived by varying the renormalization and factorization scales independently by a factor of two around their central value $Q$. 
The PDF uncertainty is evaluated according to the symmetric Hessian prescription \cite{PDF4LHC:2022cjn},

Fig.~\ref{fig:scales} presents the dependence of the inclusive cross section on the renormalization and factorization scales with $\mu_r=\mu_f=\mu$ at the 14 TeV LHC. 
As shown by the LP predictions in the top panel, including higher-order corrections significantly reduces the scale dependence.
We also observe that the NLO LP prediction is very close to the full NLO result when $\mu>Q$.
The two even coincide if the scale is chosen around $3Q$, indicating that higher-order corrections are dominated by the LP approximation.
Furthermore, we find that the full NLO result exhibits a mild sensitivity to the scale variations.
This is because the subleading power correction, albeit small in magnitude,  has a strong scale dependence which varies precisely in the opposite direction to that at LP.
The bottom panel shows the approximate higher-order predictions, including subleading power corrections at NLO. 
The higher-order effects become more pronounced as the scale increases.
The scale uncertainties would be further reduced after computing the full N$^2$LO QCD corrections.

In Tab.~\ref{tab:fixorder}, we present numerical results of the total cross sections and their scale and PDF uncertainties at different collider energies. 
At the 13 TeV LHC, the NLO correction increases the cross section by 34\%.
The aN$^2$LO result surpasses the NLO prediction by 12\%.
It is noteworthy that the scale-independent parts of the hard and soft functions contribute one-third of the $\mathcal{O}(\alpha_s^2)$ corrections. More specifically, $24\%$ ($10\%$) comes from the hard (soft) function.
The aN$^3$LO correction provides a further improvement of 2\% with respect to the aN$^2$LO result.
The aN$^3$LO prediction lies nearly within the scale uncertainty band of the aN$^2$LO result,
exhibiting an excellent perturbative convergence.
We also investigate the effect of the triple correlation in the three-loop soft function, i.e., the last term in Eq. (\ref{eq:softano}), finding that they give a $4\%$ contribution to the $\mathcal{O}(\alpha_s^3)$ correction.
From this table, one can observe that the PDF uncertainties are around 2\%, independent of the perturbative orders.
This is because we have used the same PDF set in the calculations at different orders.
The results at other collider energies display similar features.

In Tab.  \ref{tab:comparison}, we compare our predictions with the result in Refs. \cite{Kidonakis:2021vob,Kidonakis:2024kts}.
It can be seen that the two calculations agree well across various collider energies.
However, one should note that they
have used different resummation formalisms, kinematics, and scales, which can cause independent sizable discrepancies in addition to the perturbative higher-order hard, soft functions and anomalous dimensions.

\begin{table}[h]
    \centering
    \begin{tabular}{|c|c|c|} \hline
        $\sigma$ (pb) & \cite{Kidonakis:2021vob,Kidonakis:2024kts} & aN$^3$LO \\ \hline
        7 TeV & 17.2$^{+0.4+0.7}_{-0.3-0.4}$ & 17.2$^{+0.8+0.6}_{-0.9-0.4}$  \\\hline
        8 TeV & 24.6$^{+0.6+0.9}_{-0.5-0.6}$ & 24.6$^{+1.1+0.8}_{-1.3-0.5}$  \\\hline
        13 TeV  & 79.5$^{+1.9+2.0}_{-1.8-1.4}$ & 79.4$^{+3.5+1.9}_{-3.8-1.4}$ \\\hline
        13.6 TeV & 87.6$^{+2.0+2.1}_{-1.9-1.5}$ & 87.8$^{+3.9+2.1}_{-4.1-1.5}$  \\\hline
        14 TeV & 94.0$^{+2.2+2.2}_{-2.1-1.6}$ & 93.7$^{+4.1+2.1}_{-4.4-1.6}$  \\\hline
    \end{tabular}
    \caption{Comparison between our predictions and the results in \cite{Kidonakis:2021vob,Kidonakis:2024kts} with $m_t=172.5$ GeV and MSHT20 NNLO PDF. The first and second errors of our results represent the scale and PDF uncertainties, respectively.}
    \label{tab:comparison}
\end{table}

\begin{figure}
    \centering
    \includegraphics[width=0.95\linewidth]{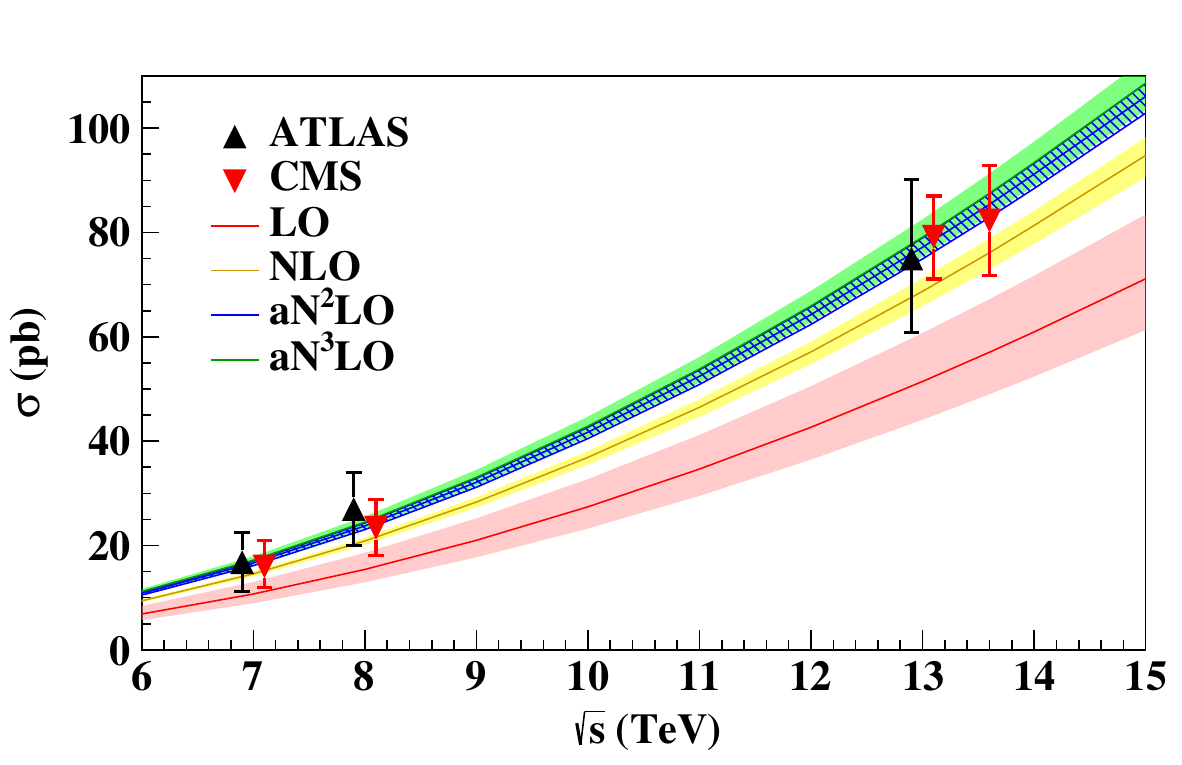}
    \caption{Comparison between theoretical predictions and experimental data.
    The bands in theoretical predictions denote the scale uncertainties.}
    \label{fig:data}
\end{figure}

Fig.~\ref{fig:data} shows
the comparison between theoretical predictions and ATLAS and CMS measurements at $\sqrt{s}$=7~\cite{ATLAS:2012bqt,CMS:2012pxd}, 8~\cite{ATLAS:2020cwj,CMS:2014fut}, 13~\cite{ATLAS:2024ppp,CMS:2022ytw}, and 13.6~\cite{CMS:2024okz}~TeV.
It can be clearly seen that the central values of the experimental data agree better with the aN$^2$LO and aN$^3$LO predictions than the NLO ones, 
underlining the necessity of including higher-order corrections.
From this comparison, we can derive the CKM matrix element $V_{tb}$ directly as the cross section is proportional to $|V_{tb}|^2$,  
given that the processes induced by a strange or down quark, $gs(gd)\to tW$,
 are highly suppressed due to the tiny values of the CKM matrix elements $V_{ts}$ ($\sim 4\times 10^{-2}$) and $V_{td}$ ($\sim 9\times 10^{-3}$) \cite{ParticleDataGroup:2024cfk}.
Consequently, we obtain $|V_{tb}|=0.99\pm 0.03({\rm expt})\pm 0.03({\rm theor})$, where the first and second errors represent the experimental and theoretical uncertainties, respectively.
This result, based on measurements of a single process, agrees with the world average value in \cite{ParticleDataGroup:2024cfk} with comparable accuracy.

\begin{figure}[t]
    \centering
    \includegraphics[width=0.95\linewidth]{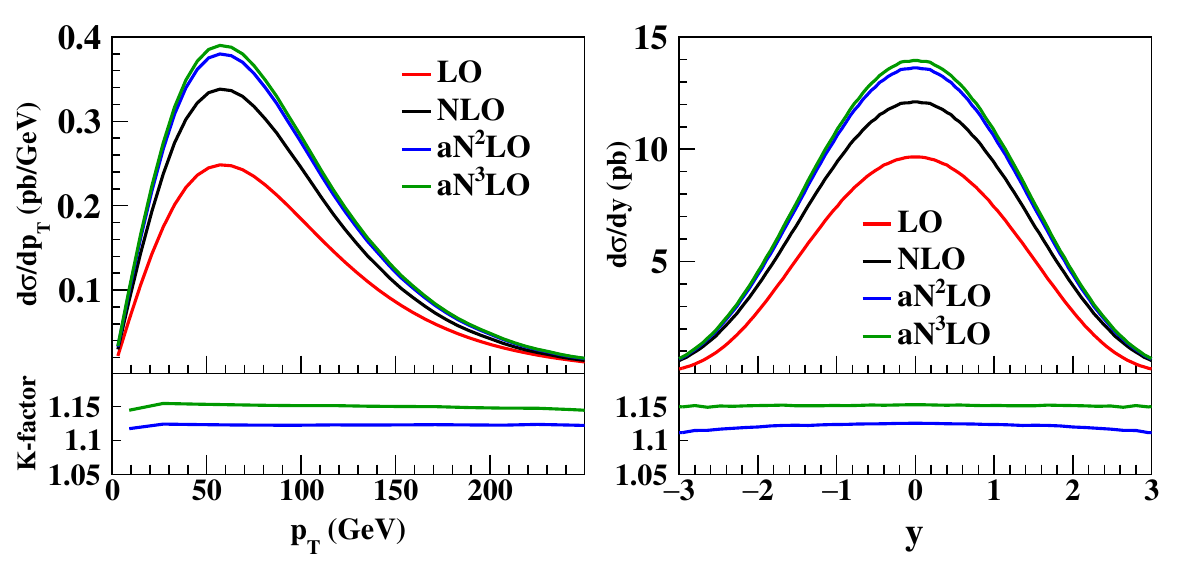}
    \caption{Top-quark $p_T$ distribution (left) and rapidity distribution (right) at the 14 TeV LHC. The K-factor is defined as the ratio of the aN$^2$LO (aN$^3$LO) over the NLO results.}
    \label{fig:distpt}
\end{figure}

Following Eq.~(\ref{eq:dsigma}), the higher-order QCD corrections to the top-quark transverse momentum and rapidity distributions can be derived straightforwardly in the threshold limit.
The resulting distributions are shown in Fig.~\ref{fig:distpt}.
The aN$^2$LO and aN$^3$LO corrections are sizable across the full kinematic regions, and thus should be included in the precise predictions.
Since the enhancements exhibit only weak kinematic dependence, 
it is sufficient to apply a universal K-factor that is defined as the ratio of higher-order over NLO predictions.

In summary, we have calculated the approximate N$^2$LO and N$^3$LO QCD corrections for $tW$ associated production at the LHC, which constitute the dominant contributions to the corresponding full perturbative predictions. Our calculation incorporates the N$^2$LO hard and soft functions as well as the complete N$^3$LO scale-dependent contributions. 
The approximate N$^2$LO and N$^3$LO corrections increase the NLO cross section by more than $10\%$. 
Comparison with the experimental data
allows us to determine the CKM matrix element $|V_{tb}| = 0.99 \pm 0.03 (\text{expt}) \pm 0.03 (\text{theor})$ directly without assuming unitarity,
which is comparable with the current world average value. 
Furthermore, these higher-order corrections have sizable impacts on kinematic distributions. 
Collectively, our results 
provide valuable theoretical input for ongoing and future experimental analyses, given that $tW$ production is one of the main backgrounds at the LHC, e.g., in the search for the Higgs pair signal \cite{ATLAS:2024pov}.

Looking forward, a complete N$^2$LO calculation will be essential to further reduce theoretical uncertainties. Achieving this goal requires the development of a robust subtraction scheme for the top-quark-pair interference contribution in the $tW\bar{b}g$ final state, along with a coherent combination of the double-virtual, virtual-real, and double-real corrections.

We would like to thank Long-Bin Chen, Liang Dong, Zhao Li and Yefan Wang for collaboration on the computation of the two-loop hard function.
We thank Wenbin Qian for useful discussion on the statistical analysis.
This work was partly supported by the National Natural Science Foundation of China under Grants No. 12275156 and No. 12321005, the Taishan Scholar Foundation of Shandong province (tsqn201909011), and the Shandong Provincial Department of Science and Technology under Project No. TQ012025001. 
The Feynman diagrams were drawn using {\tt TikZ-FeynHand} \cite{Dohse:2018vqo}. The authors gratefully acknowledge the valuable discussions and insights provided by the members of the Collaboration of Precision Testing and New Physics.

\bibliographystyle{apsrev4-1}
\bibliography{bib}

\end{document}